\documentclass[preprint,floatfix,amsmath,amssymb,prb]{revtex4}
\usepackage{graphicx,float,amssymb,subfigure}
\usepackage{dcolumn}
\begin{document}
\title{Two dimensional semiconductors with possible high room temperature mobility}
\author{Wenxu Zhang}\email{xwzhang@uestc.edu.cn}
\author{Zhishuo Huang, Wanli Zhang}

\affiliation{State key laboratory of electronic thin films and integrated devices, University of
Electronic Science and Technology of China, Chengdu 610054, China}
\begin{abstract}
We calculated the longitudinal acoustic phonon limited electron mobility of 14 two dimensional semiconductors with composition of MX$_2$, where M (= Mo, W, Sn, Hf, Zr and Pt) is the transition metal, and X is S, Se and Te. We treated the scattering matrix by deformation potential approximation. We found that out of the 14 compounds, MoTe$_2$, HfSe$_2$ and HfTe$_2$, are promising regarding to the possible high mobility and finite band gap. The phonon limited mobility can be above 2500 cm$^2$V$^{-1}$s$^{-1}$ at room temperature.
\end{abstract}
\maketitle
\section{Introduction}
Two dimensional (2D) crystals are materials with thickness of several atomic layers and extended periodically in the other two dimensions. Examples of them are monolayer or multilayer of graphene, MoS$_2$, BN, MoO$_3$ and so on. These 2D materials have now received a lot of research interests because of its unique physical properties, such as the Dirac cone in electronic band structures in graphene. Applications of these new materials have been demonstrated. High speed radio frequency devices are fabricated which make full use of the ultrahigh electron mobility in graphene. The transition-metal dichalcogenide semiconductor MoS$_2$ has also attracted great interest because of its distinctive electronic, optical and catalytic properties, as well as its importance for dry lubrication. Logical devices were also fabricated based on MoS$_2$ 2D crystals. Field effect transistors based on 2D materials are promising because it can overcome the shorted-channel effect, which is one of the biggest obstacles of further decrease of the dimensions of semiconductors\cite{Liu2012}. Thus 2D semiconductors are attractive for semiconductor technology after silicon.
\par Regarding to the applications of 2D materials in logical devices of semiconductors, the present materials are not good enough. Graphene has ultrahigh mobility, but it is intrinsically metallic. It is possible to open a gap at the Dirac cone, but usually the gap is tiny, and required extra parameters, for example, external electric fields, which is not easy to fully integrate it with present semiconductor processes. Monolayer MoS$_2$ has a direct band gap about 1.8 eV at room temperature. Transistors fabricated on 5 nm thick MoS$_2$ show no short channel effects down to a channel length of $\sim$100 nm without aggressive gate oxide scaling\cite{Liu2012}. But the mobility was shown to be too low for practical applications at room temperature. The mobility can be enhanced by encapsulating monolayer MoS$_2$ in a high-$k$ dielectric environment\cite{Radisavljevic2013}. On the device level, the mobility can be further engineered by electron doping.  Other factors such as electrode materials are crucial to the the measured mobility\cite{popov12}. Thus, it is important to know the intrinsic limited mobility in order to further improve the performances of the devices. The theoretical limits of these mobility is low as calculated by Kaasbjerg\cite{kaasbjerg2012} and Li\cite{Li2013}. Phonon is one of the most important scattering sources of electron transport. Phonon limited mobility in MoS$_2$ was carefully investigated by Kaasbjerg \emph{et al} \cite{kaasbjerg2012, kaasbjerg2013}. The contributions from acoustic and optical phonon are included and electron-phonon coupling matrices are calculated by frozen phonon methods. Electron-phonon, as well as piezoelectric interactions, are taken into account. The calculated room temperature mobility is about 410 cm$^2$V$^{-1}$s$^{-1}$ which sets the upper bound of intrinsic mobility. And also according to the work of Yoon\cite{Yoon2011}, due to the heavier electron  effective mass and a lower mobility, MoS$_2$ is not ideal for high-performance applications. Is it possible to find other 2D semiconductors with suitable band gap and higher mobilities? High through output calculations based on density functional theory (DFT) has been shown to be a fast way to screen out materials with desirable properties if suitable descriptors are invented. Ciraci \emph{et al} \cite{ataca12} predicted 52 different stable MX$_2$ single layer compounds from 88 different combinations. Leb\`{e}gue \emph{et al} \cite{Lebegue2013} use data filtering and \emph{ab initio} calculations and screen 92 2D compounds out of the whole ICSD database. In this work, we performed electronic calculations of the selected semiconductors with composition of MX$_2$, where M (= Mo W, Sn, Hf, Zr or Pt) is the transitional metal, and X is S, Se or Te. In order to fast screen out the materials with high performances, mobility limited by long wave acoustic phonon was estimated by calculating the deformation potential. We found that out of the 14 compounds, three compounds, MoTe$_2$, HfSe$_2$ and HfTe$_2$, are promising regarding to their mobility and band gap. The phonon limited mobility can be above 2500 cm$^2$V$^{-1}$s$^{-1}$.
\section{Calculation details}
The calculations were performed by the
full-potential local-orbital code\cite{fplo} in the version
FPLO9.00-33 with the default basis settings. All calculations
were done within the scalar relativistic approximation. The local density approximation
functional was chosen to be that parameterized by Perdew and Wang.\cite{gga}
The number of k-points in the whole Brillouin zone was set to $32\times32\times5$
in order to ensure the convergency of the results. Convergency of the total energy was set to be better than $10^{-8}$ Hartree together with that of the electron density better than $10^{-6}$ in the internal unit of the codes. Supercell with vacuum layer of 30 \AA\ was used to model the 2D nature of the compounds within the 3D crystal cell. Within the deformation potential approximation\cite{Bardeen1950,takagi96}, the electron mobility (Takagi model) is approximated by
\begin{equation}
\mu=\frac{e\hbar^3\rho V^2_s}{k_B T m_e m_d E^2_{el-ph}}=\frac{e \hbar^3c_{11}}{k_BTm_em_dE^2_{el-ph}}\label{equ:mobility}
\end{equation}
where $k_B$ and $\hbar$ are the Boltzmann constant and the Planck constant, respectively. $m_e$ is the effective mass of electron and $m_d$ is the electron density of state mass. $\rho$ is the density of the material and V$_s$ is the sound velocity in the corresponding direction. Li \emph{et al} \cite{Li2013} calculated the electrical transport of monolayer MoS$_2$ by the linear response method. It shows that the longitudinal phonon have the strongest interaction with electrons. Bruzzone and Fiori used this Takagi model to compute the electron mobility of hydrogenated and fluorinated graphene as well as h-BCN from first principles\cite{Bruzzone2011} and show that graphene with a reduced degree of hydrogenation can compete with silicon technology. The sound velocity (V$_s$) is calculated by the supercell method, where frozen phonon mode corresponding to the longitudinal phonon with vector $\mathbf{q}=\frac{\pi}{8a}(1,0,0)$ was simulated. The phonon frequency ($\omega_k$) was obtained, which is related to the sound velocity by $\omega_k=V_s q$. The elastic constants $c_{11}$ in hexagonal crystal is related to the sound velocity by $c_{11}=\rho V_s^2$. The electron phonon coupling (E$_{el-ph}$) was approximated by the deformation potential. Crosschecking of the calculated parameters was done with pseudo-potential code Quantum Espresso\cite{QE}.

\section{Long wave acoustic phonon limited mobility in MX$_2$}
\begin{figure}
\includegraphics[scale=0.45]{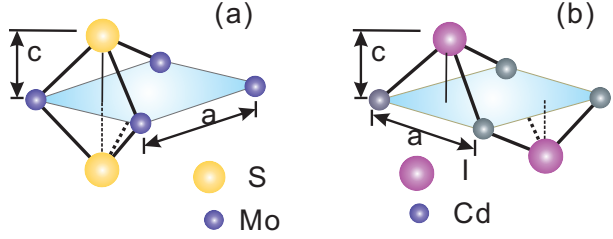}
\caption{\label{fig:str} Schematic illustration of the crystal structure of MoS$_2$ and CdI$_2$. The lattice parameters are $a$ and $c$.}
\end{figure}

Out of the ICSD database, only 16 compounds with MX$_2$ are semiconductors\cite{Lebegue2013}.  There are more layered compounds with more chemical elements. The complexity may hinder its applications. We selected 14 MX$_2$ compounds which crystallizes into two different crystal structures. One is MoS$_2$ and the other is CdI$_2$. The difference is that the anion hexagonal nets are A-A stacked in MoS$_2$, while they are A-B stacked in CdI$_2$ as shown in Fig.\ref{fig:str}. The phonon limited mobility for the 14 semiconductors are listed in Tab. \ref{table:t1}.
\begin{figure}
\includegraphics[scale=0.45]{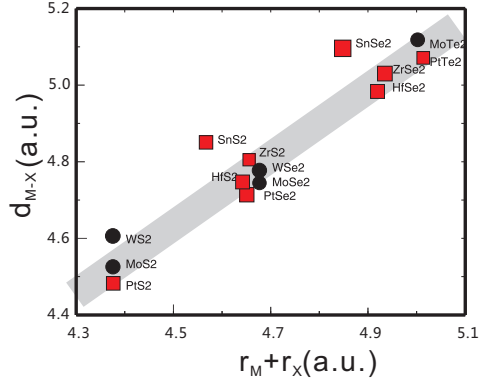}
\caption{\label{fig:lr} Distances of M and X vs. covalent radii of the components (r$_M$+r$_X$). The shadowed area is for guiding eyes.}
\end{figure}
The other parameters which determined the mobility according to Equ.(\ref{equ:mobility}) are also listed in the table. The distance between M and X are $d_{M-X}=\sqrt{(a^2/3+c^2)}$. We plotted $d_{M-X}$ as a function of the summation of atomic covalent radii of M and X (r$_M$+r$_X$) as shown in Fig.\ref{fig:lr}. The data show a good linear proportionality. This is in accordance with the covalent bonding of the material. The sound velocity in MoS$_2$ is 7.93 km/s, which is the highest value among the compounds, while that of HfSe$_2$ with 4.72 km/s is the lowest. This leads to the relatively high elastic constant $c_{11}$ in MoS$_2$ and low value in HfSe$_2$. Among the compounds, MoS$_2$, WS$_2$ and PtS$_2$ have large $c_{11}$'s, but their mobilities are not large. The differences of the mobility among the compounds are mostly within 50\%.
\par The electron effective mass is almost isotropic in the compounds with MoS$_2$ structure, while it is quite anisotropic in the compounds with CdI$_2$ structure. The difference can be as large as ten folds, but the three compounds with Pt is an exception, where the effective mass is almost isotropic. The difference in the effective mass leads to anisotropic transport properties as studied in bulk materials of HfS$_2$.
\par The mobility of MoS$_2$ were calculated by Li\cite{Li2013}, where a full treatment of scattering by phonon was  done. The electron phonon interaction matrices were calculated in the perturbation way. It can be seen that in the case of MoS$_2$, the acoustic phonon limited mobilities at 300 K obtained by the two methods are quite comparable. It is 320 cm$^2$V$^{-1}$s$^{-1}$ if the full contributions from phonon were included, while it is 340 cm$^2$V$^{-1}$s$^{-1}$ in our results. The sound velocity of longitudinal acoustic phonon and intra valley deformation potential extracted by Li are 6.6 km/s and 3.20 eV, comparable with our results of 7.93 km/s and 3.90 eV.
\begin{table}
  \centering
  \caption{Structural, mechanical and electronic parameters of the semiconductors calculated by LDA. The effective mass in the $\Gamma-K$ direction for the MoS$_2$ structure and $\Gamma-M$ direction for the CdI$_2$ structure is calculated.}\label{table:t1}
  \begin{tabular}{lcccccccc}
  \hline
 MX$_2$ &a    & c    & V$_s$ & c$_{11}$&m$^\ast_{\Gamma-K(M)}$ & m$^\ast_{K-M}$ &E$_{el-ph}$ & $\mu$ \\
      & (a.u.)&(a.u.)& (km/s)  &($10^{11}N/m^{2}$)&(m$_e$)&(m$_e$)&(eV)&(cm$^2$/V$\cdot$ s)\\
  \hline
  MoS$_2$  & 5.927 & 2.962 & 7.93 & 6.25& 0.45 & 0.45 & 3.90 & 340  \\
  MoSe$_2$ & 6.168 & 3.156 & 6.01 & 4.94& 0.52 & 0.52 & 3.65 & 240  \\
  MoTe$_2$ & 6.618 & 3.411 & 5.04 & 4.45& 0.53 & 0.57 & 0.92 & 2526 \\
  WS$_2$   & 6.047 & 2.992 & 6.67 & 6.52& 0.24 & 0.26 & 3.92 & 1103 \\
  WSe$_2$  & 6.166 & 3.164 & 5.55 & 5.66& 0.33 & 0.31 & 3.78 & 705  \\
  \hline
  SnS$_2$  & 6.879 & 2.797 & 6.18 & 3.42& 2.11 & 0.21 & 3.55 & 306 \\
  SnSe$_2$ & 7.165 & 2.999 & 4.83 & 2.71& 2.91 & 0.17 & 2.91 & 447 \\
  HfS$_2$  & 6.731 & 2.750 & 5.86 & 4.33& 3.30 & 0.24 & 1.31 & 1833 \\
  HfSe$_2$ & 6.944 & 2.978 & 4.72 & 3.37& 3.10 & 0.18 & 1.08 & 3579 \\
  ZrS$_2$  & 6.817 & 2.771 & 7.21 & 4.05& 1.62 & 0.31 & 1.52 & 1247 \\
  ZrSe$_2$ & 7.007 & 3.008 & 5.42 & 3.18& 2.03 & 0.22 & 1.25 & 2316 \\
  PtS$_2$  & 6.670 & 2.327 & 6.61 & 7.05& 0.26 & 0.25 & 3.63 & 1107 \\
  PtSe$_2$ & 6.978 & 2.464 & 4.73 & 4.26& 0.21 & 0.19 & 2.86 & 1892 \\
  PtTe$_2$ & 7.485 & 2.634 & 4.89 & 4.72& 0.90 & 0.77 & 1.73 & 367  \\
   \hline
  \end{tabular}

\end{table}

\par
\begin{figure}
\includegraphics[scale=0.45]{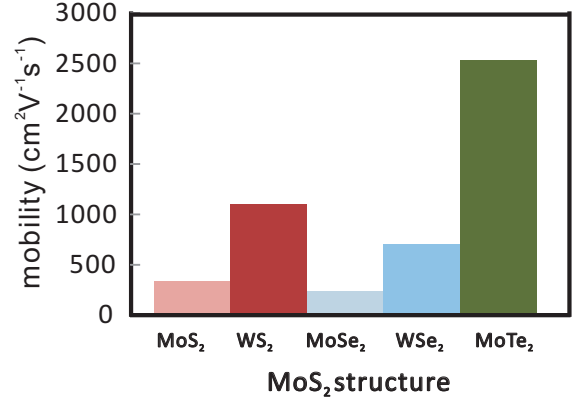}
\caption{\label{fig:mob-h} Electronic mobility of compounds with MoS$_2$ structure.}
\end{figure}

\begin{figure}
\includegraphics[scale=0.45]{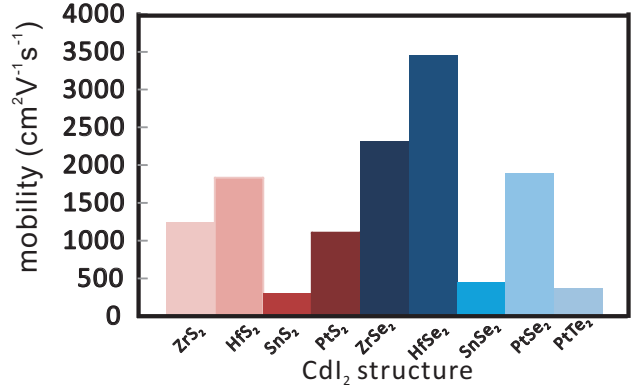}
\caption{\label{fig:mob-t} Electronic mobility of compounds with CdI$_2$ structure.}
\end{figure}

 The electron phonon coupling constant and the electron effective mass in the denominator play the most important role in determining the electron mobility. The mobility and the electron phonon coupling constant are plotted in Fig. \ref{fig:mob-h} and \ref{fig:mob-t}. Combined with the small effective mass, MoTe$_2$, HfSe$_2$ and ZrSe$_2$ show a large upper bound of acoustic limited mobility. The values are even ten times larger than the well studied MoS$_2$.

\section{Electronic structures}
\begin{figure}
\subfigure{\label{fig:sub-mote2}
\includegraphics[scale=0.45]{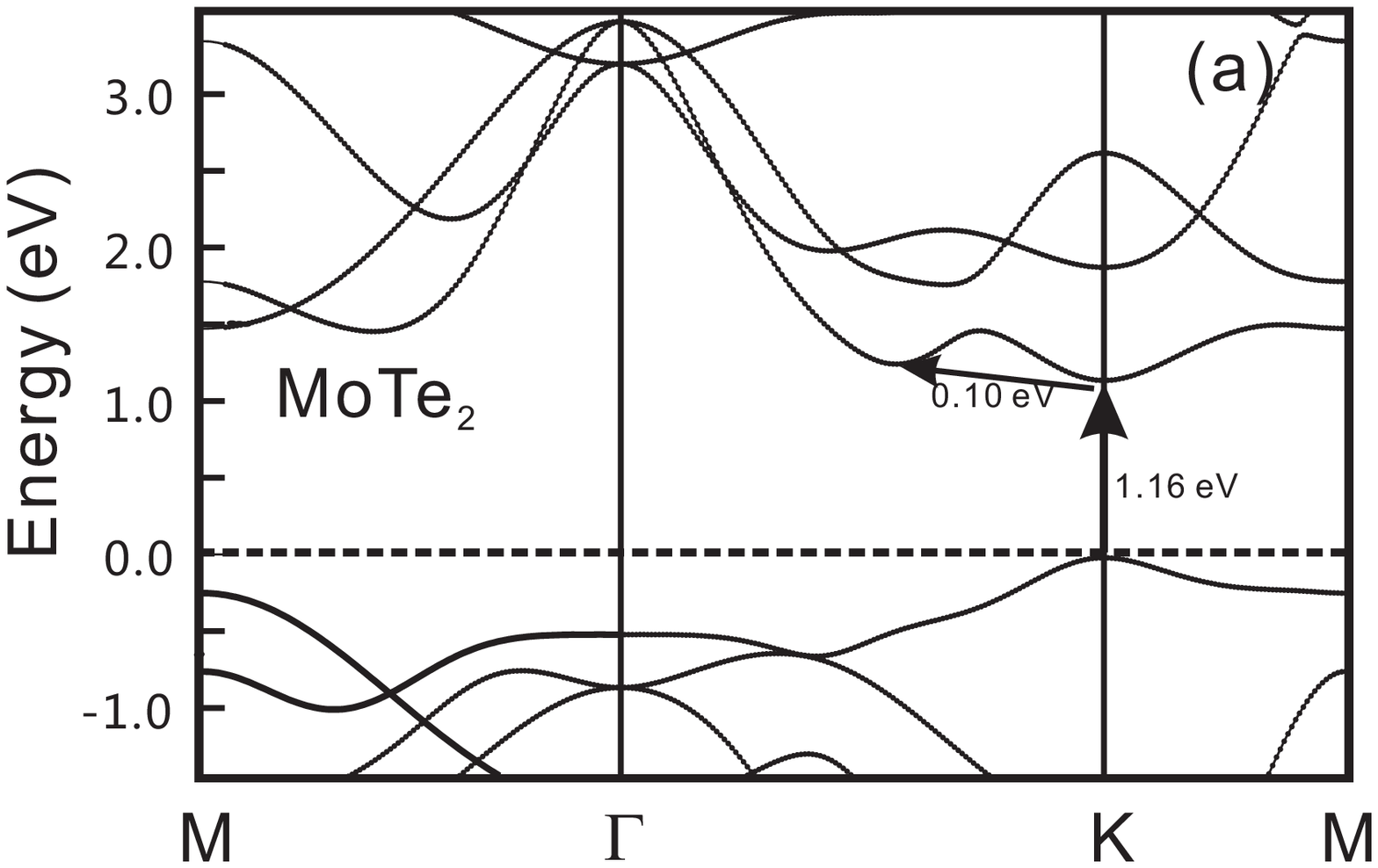}}
\subfigure{\label{fig:sub-mos2}
\includegraphics[scale=0.45]{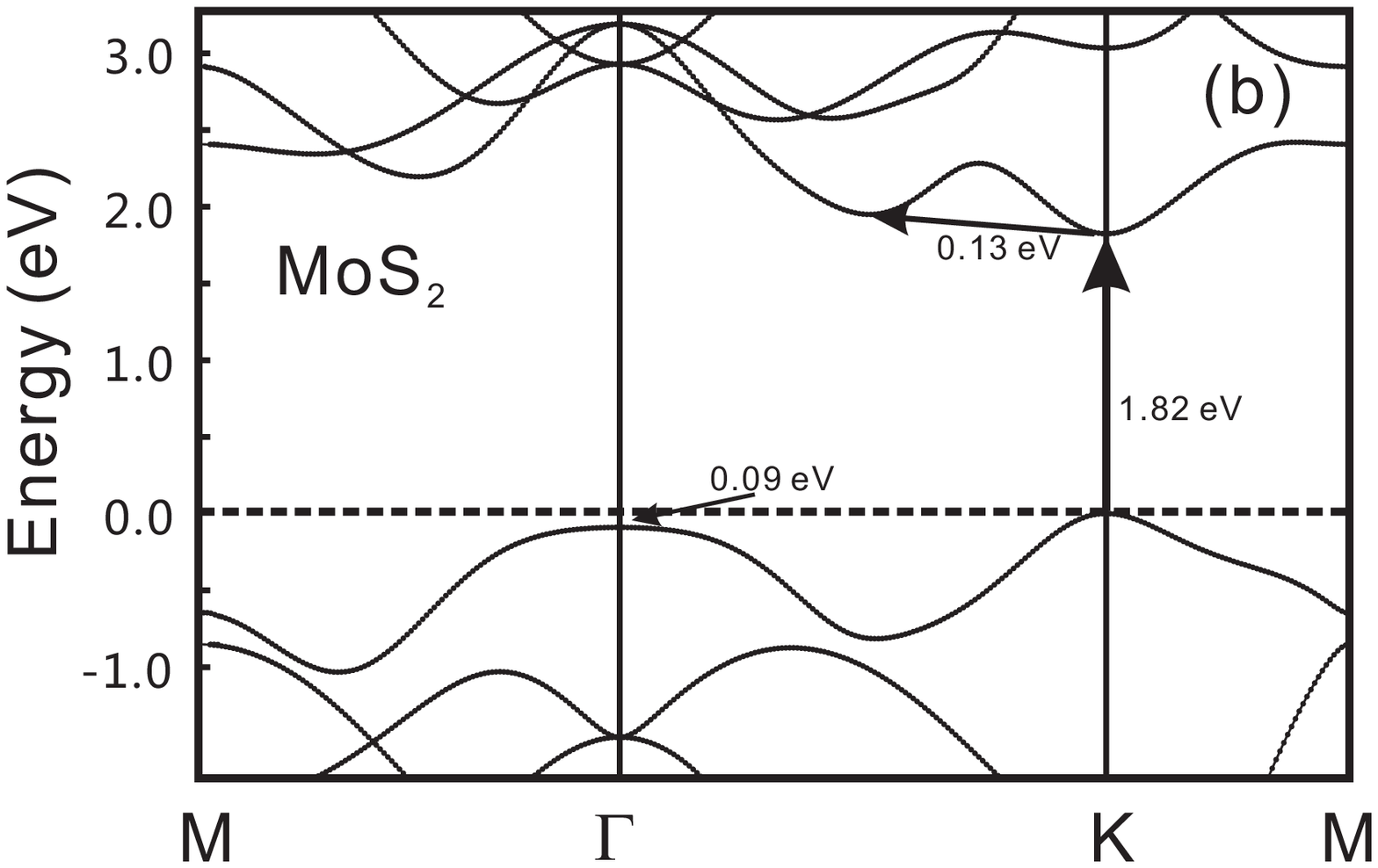}}
\caption{\label{fig:mos2bands} Band structure of MoTe$_2$(a) and MoS$_2$(b). The arrows show the energy differences between the extremals.}
\end{figure}

We have seen that these compounds have two different structures. The MoS$_2$ structure and CdI$_2$ structure which are different only by a relative shift of the anion hexagonal. This structural difference leads to different electronic structure. We show only the electronic bands of the compounds with possible high mobility. The bands of MoTe$_2$ are shown in Fig.\ref{fig:sub-mote2}, while those of MoS$_2$ which are extensively studied in literatures are shown in Fig.\ref{fig:sub-mos2}. Both of them are direct band gap semiconductors. The valence band maximum (VBM) and conduction band minimum (CBM) are located at the $K$-point. The direct band gaps in these compounds are 1.16 eV (MoTe$_2$) and 1.82 eV (MoS$_2$), respectively. There is a second local minimum along the $\Gamma-K$ direction. These local minima are 0.13 eV and 0.10 eV above the CBM in MoS$_2$ and MoTe$_2$ respectively. This is considered as a possible scattering states for electrons as discussed by Li\cite{Li2013}. It can dramatically decrease the electron mobility because of the increased scattering rate.  The VBM in MoS$_2$ is only 0.09 eV above the local maximum at the $\Gamma$-point. However it is a quite heave hole as indicated by the small curvature of the bands compared with that at K-point.
 \par
\begin{figure}
\subfigure{\label{fig:sub-zrse2}
\includegraphics[scale=0.4]{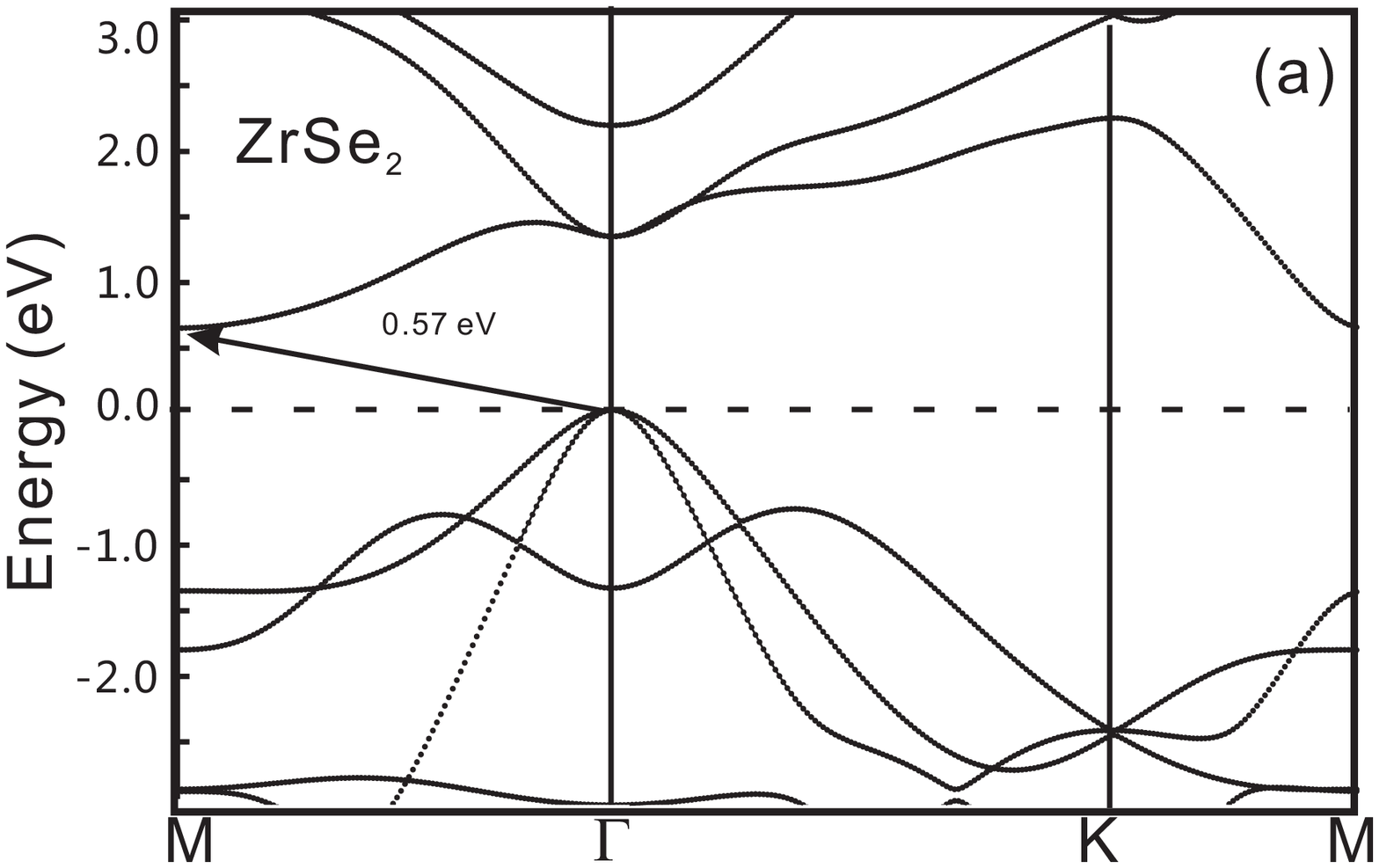}}
\subfigure{\label{fig:sub-hfse2}
\includegraphics[scale=0.4]{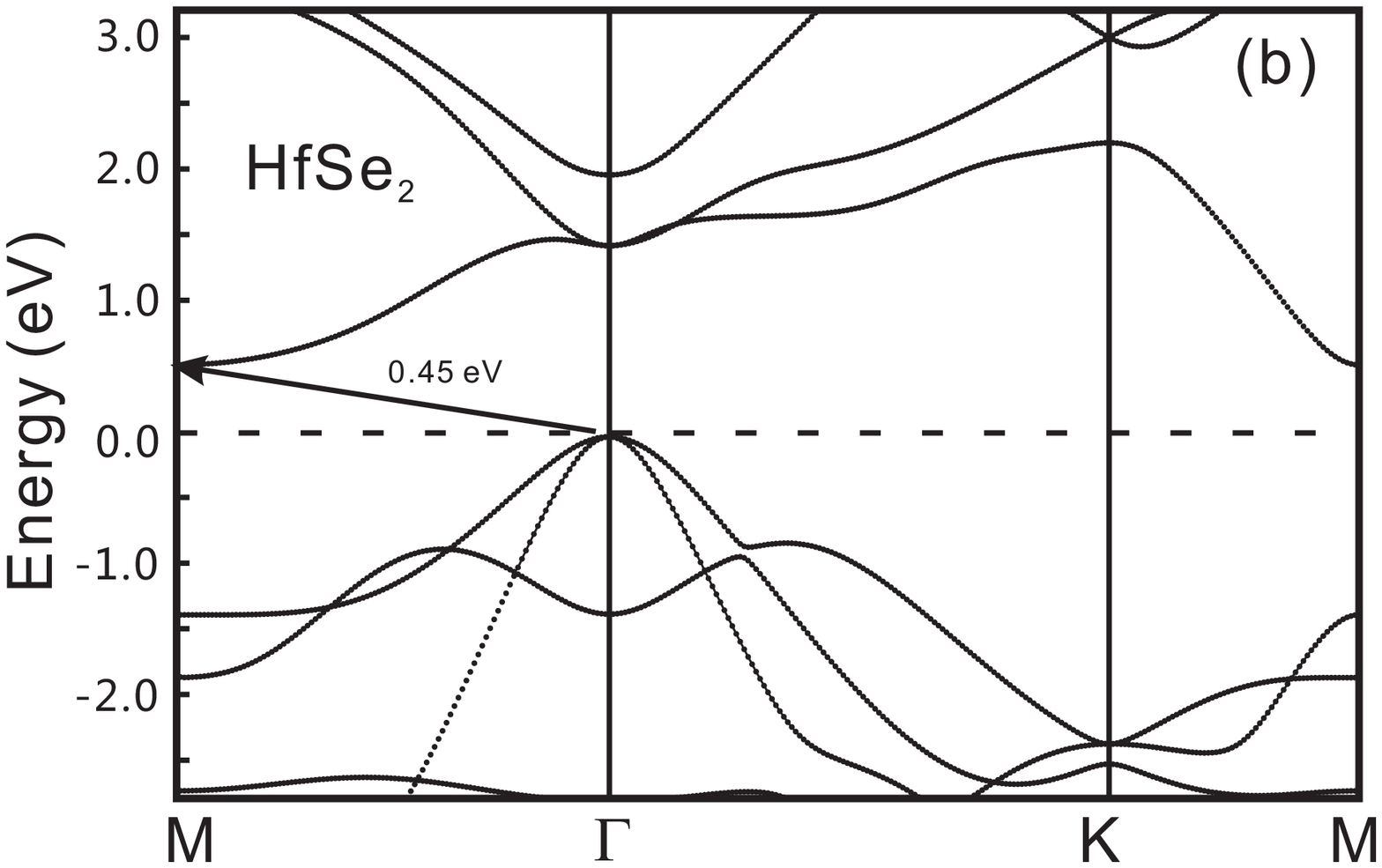}}
\caption{\label{fig:cdi2bands} Band structure of ZrSe$_2$(a) and HfSe$_2$(b).The arrows show the energy differences between the extremals.}
\end{figure}

 The bands of HfSe$_2$ and ZrSe$_2$ are different from those of MoTe$_2$, because of different crystal symmetry and atomic bonding. But there is a great similarity of the bands of the two compounds as shown in Fig.\ref{fig:cdi2bands}. It can been seen that the dispersions around the M-point of the compounds with MoS$_2$ structure are quite isotropic while those with the CdI$_2$ structure are anisotropic. This leads to the different electron masses along the different directions as already shown in Table \ref{table:t1}. The effective mass in the $\Gamma-M$ direction is about ten times larger than that in the $M-K$ direction. Compounds with the MoS$_2$ structure are direct gap semiconductors while those with the Cd$_2$ structure have indirect gaps. The indirect LDA bandgap is between the $\Gamma$-point and $M$-point. The bandgap $E_g$ are 0.57 eV and 0.45 eV for ZrSe$_2$ and HfSe$_2$, as shown in Figs. \ref{fig:sub-zrse2},\ref{fig:sub-hfse2} respectively.

\section{Discussions}
In this work, we have considered only the long wave acoustic phonon scattering. Of course, there are many other scattering mechanisms which limits the mobility. As the Matthiessen rule, i.e., $\frac{1}{\mu_{total}}=\frac{1}{\mu_{phonon}}+\frac{1}{\mu_{impurity}}+\frac{1}{\mu_{electron}}+\cdots$, where the scattering sources are phonons, impurities, electrons and so on, may hold, the mobility will be limited by any one of these mechanisms. There are uncertainties to precisely determine the contributions from each of these mechanism, both experimentally and theoretically, so that we cannot predict the mobilities precisely. However, by computing selected one of them, the upper bound of the mobility can be set. We thus can say, it is hopefully that we can find compounds with possible high mobility among the selected ones with larger upper bounds. More sophisticated calculations are needed to aim more precisely. However, these calculations are time consuming for search within the large amount of candidates.
\par
According to our estimation, the mobility of WSe$_2$ may be larger than MoS$_2$. Mobilities of WSe$_2$ and MoS$_2$ are extracted from the transfer character curves of field-effect transistors \cite{Fang2013}. It is shown that the electron mobility in WSe$_2$ is about 110 cm$^2$V$^{-1}$s$^{-1}$, while that of MoS$_2$ is about 25 cm$^2$V$^{-1}$s$^{-1}$. These experimental results can be an example of our prediction.
\section{Conclusion}
In this work, the electron mobility of 14 MX$_2$ type two dimensional semiconductors were calculated where only elastic scattering from long wave acoustic phonon was taken into account by the deformation potential approximation. We found that the mobility of the semiconductors with CdI$_2$ structures are generally larger than that of MoS$_2$ structure. However, the electronic bands are anisotropic with the CdI$_2$, which means that their electronic transport properties are dependent on directions. MoTe$_2$, ZrSe$_2$ and HfSe$_2$ are more promising two dimensional semiconductor than MoS$_2$ when considering their possible larger carrier mobilities and sizeable band gap. The acoustic phonon limited electron mobility are above 2000 cm$^2$V$^{-1}$s$^{-1}$ at room temperature.
\section{Acknowledgement}
Financial support from the Research Grant of Chinese Universities and International Science \& Technology Cooperation Program of China (2012DFA51430) are acknowledged.

\end{document}